\documentclass[a4, 11pt]{article}          
\usepackage[textwidth=16cm]{geometry}
\usepackage{amssymb,amsmath,amsfonts}
\usepackage{graphicx}
\usepackage{color,bm,cite,authblk}

\newcommand{\fig}[1]{Fig.~\ref{#1}}
\newcommand{\eq}[1]{Eq.~(\ref{#1})}

\newcommand{\etal}{{\it et. al.}}

\begin{document}
\title{Kinetic Monte Carlo simulations of self organized nanostructures on Ta Surface Fabricated by Low Energy Ion Sputtering}
\author[1]{Shalik Ram Joshi}
\author[2]{Trilochan Bagarti} 
\author[1]{Shikha Varma\thanks{shikha@iopb.res.in}}

\affil[1]{Institute of Physics, Sachivalaya Marg, Bhubaneswar-751005}
\affil[2]{Harish-Chandra Research Institute\\Chhatnag Road, Jhunsi, Allahabad-211019}
\maketitle

\begin{abstract}
Surfaces bombarded with low energy ion beams often display development of self assembled 
patterns and quasi-periodic structures. Kinetic Monte Carlo simulations have been performed 
to describe ion sputtered Tantalum surfaces. A weak nonlinearity in the relaxation process 
has been introduced and the results show that the Positive Schwoebel barrier, produced by the 
nonlinear Hamiltonian, is necessary in describing ion bombarded Tantalum surfaces. Furthermore,
their scaling exponents suggest presence of a class other than KPZ. 
\end{abstract}

\section{Introduction}
\label{introduction} 
Non-equilibrium surfaces produced via bombardment with energetic ion beams, often exhibit 
well ordered patterns having several potential applications \cite{Qia,Eli,Mar}. 
The surface morphology, here, develops as a consequence of competition between 
a variety of processes like roughening dynamics, relaxation processes, generation of defects, 
material transport etc. \cite{Mak,Fro}. Occurrence of a wide array of surface morphologies 
complicates the prediction of dominant mechanism controlling their evolution. Non-metallic 
surfaces generally display hills and depressions for normal incident ion beams 
\cite{Fac}, while showing quasi-periodic ripple morphology under off-normal conditions \cite{Cha,Car}. 
For metallic surfaces, ripples develop even at normal incidence \cite{Rus}. 
Symmetry breaking anisotropy in surface diffusion can cause these ripples to rotate with substrate 
temperature \cite{Rus,Val}. For low energy ion beam induced patterning, although the erosive 
processes are dominant, enhanced surface diffusion due to the defect mobility also becomes important 
especially under the low ion-flux conditions \cite{Mak,Cha}. The surface morphology in this 
scenario is governed by the non-equilibrium biased diffusion current.

Continuum model approach has proved to be a very successful technique in describing the surface 
evolution but involvement of several complex phenomena make it difficult to relate with the 
experimental results \cite{Cha1}. An alternative approach is the Kinetic Monte Carlo(KMC) method where the 
kinetic behavior of the surface is simulated at microscopic level for a discrete surface. 
During diffusion or sputtering, the surface gets modified in the units of whole atoms, only at 
the specific site \cite{Bar}. Various models have been proposed to understand the effect 
of sputtering related erosion on surface evolution. For instance, in model by Cuerno et.al 
\cite{Cue}, the local surface morphology has been used to determine the sputtering yield of 
the surface, while in binary collision approximation, the atom gets removed from the 
surface with a probability proportional to the energy deposited, in its near surface region, 
by incoming ions \cite{Bro,Har,Ste}. Surface diffusion has been formulated 
via relaxation of the surface to its minimum energy through a series of atomic jumps 
with probability that depends upon the energy of the initial and the final state \cite{Sie}. Thermally activated 
hopping, where an atom hops over an energetic barrier with a barrier height that depends on the 
local configuration, has also been considered for studying surface diffusion 
\cite{Bro,Ste,Yew}. 

In the present article, KMC results are presented in 1+1 dimension for ion beam modified 
metallic surfaces. Experimentally, the presence of ripple morphology on ion
irradiated Tantalum surfaces has been observed \cite{Ram,Sub}. We have developed a model based
on earlier work by Cuerno \etal \cite{Cue} which was not able to describe the surface
morphology of ion sputtered metallic Tantalum surface. The Schwoebel effect is found to
be important for Tantalum surface, and has been incorporated in our model, by including a
weak non-linearity in Hamiltonian for relaxation of diffusing atoms on the sputtered surface.
Simulation results, presented here, shows that the presence of Schwoebel effect produces the
surface morphology and scaling exponents that are consistent with our experimental observations.
The scaling exponents indicates that the morphology of ion sputtered Tantalum surfaces may belong
to universality class other than Edward-Wilkinson(EW) and Kardar-Parisi-Zhang(KPZ).
The paper is organized as follows. In Section \ref{model} we describe the KMC model. 
The experimental details are described in Section \ref{exp} and in Section 
\ref{result} we discuss the results.

\section{The Model}
\label{model}
In our model we assume that ion irradiation causes the surface to evolve by two dominant processes, 
namely erosion and surface diffusion. The erosion process consists of the removal of atoms 
from the surface due to the impinging ions. Ions bombarded on the surface penetrate deep into the 
substrate releasing energy to the neighboring atoms along the trajectory. If the energy gained by 
a surface atom is sufficiently large, it gets eroded from the surface. 
Sputtering yield, $Y(\phi)$, defined as the number of atoms eroded for every incoming ion at 
an angle $\phi$ to the surface normal. It depends on the energy of the ion and local morphology of the interface
and, is a measure of the efficiency of the sputtering process \cite{Ken}. 
We assume $Y(\phi) = a + b \phi^2 + c \phi^4$, where $a, b$ and $c$ are constant such that $Y(\pi/2)=0$ 
and for some critical $\phi_0$, $Y(\phi)$ has a maxima. The erosion process brings in an effective 
negative surface tension that causes the surface to become rough \cite{Bra}.

Surface diffusion, on the other hand, consists of the random migration of surface atoms on the surface 
such that the surface energy is minimized. Its strength depends on the temperature of the substrate 
and the binding energy of the atoms. The negative surface tension, during erosion, leads the system 
towards instability and as a result the system responds to restore stability by surface diffusion \cite{Cue}. 
   
We consider a one dimensional lattice with periodic boundary conditions. The surface at any instant 
of time \textit{t} is described by the height $h_{i}(t)$ at each lattice site $i=1,\ldots,L $. We 
consider initially (at time $t=0$) a flat surface $h_i(0) = \mbox{const}$. The erosion takes place 
with probability $p$ while the diffusive process occurs with probability $(1-p)$ at a randomly chosen 
lattice site. The surface is evolved by following dynamical rules.

(i) {\it Erosion:} The particle on the surface at site $i$ gets eroded with a probability $Y(\phi_i)p_e$ 
where $\phi_i = \tan^{-1}((h_{i+1}-h_{i-1})/2)$ and  $1/7 \leq p_e \leq 1$ is the ratio of the number 
of occupied neighbors to the total number of neighboring sites \cite{Ken}. The ratio $p_{e}$ accounts for the 
unstable erosion mechanism due to the finite penetration depth of the bombarding ions into the eroded 
substrate \cite{Cue,Bra,Cue1}.\\

(ii) {\it Surface diffusion:} The surface diffusion process is taken into account by nearest neighbor 
hopping. The hopping rate from an initial state $i$ to a final state $f$ is given by 
$w_{i,f} = [1+\exp(\beta \Delta H_{i\rightarrow f})]^{-1}$ where 
$\Delta H_{i\rightarrow f} = H_{f} - H_{i}$ is the difference in the energy of the states 
and $\beta^{-1} = K_{B} T$ where $T$ is the surface temperature and K$_{B}$ is the Boltzmann constant. 
The surface Hamiltonian is given by
\begin{equation}
H = \frac{J}{2} \sum_{\langle i,j \rangle} |h_i-h_j|^2,
\label{hamiltonian1}
\end{equation}
where ${\langle i,j \rangle}$ denotes sum over nearest neighbor sites and $J$ is the coupling strength. 
Similar relaxation behaviour has been considered by Cuerno \etal \cite{Cue} to describe
the surfaces evolving from initial ripple structure into rough surfaces of KPZ class. Morphology 
of several other non-metallic surfaces, post ion irradiation, have also been describe sufficiently 
well by this model \cite{Ell,Yan}.

The morphology of metallic surfaces, however, cannot be described by the above model alone. 
In these cases, a diffusing atom is repelled from the lower step and preferably 
diffuses in the uphill direction. This Schwoebel effect is controlled by the potential
barrier, Schwoebel barrier, and has been considered to be crucial for MBE grown surfaces 
\cite{Sie,Fer,Sie1}. Growth models based on MBE studies, have shown that positive Schwoebel
effect can be incorporated in the relaxation process Hamiltonian by the inclusion of a 
quartic term \cite{Sie}. In the present study, we model the relaxation behaviour on ion 
sputtered Tantalum metal surfaces by modifying \eq{hamiltonian1} to include the Schwoebel effect
in the Hamiltonian: 
\begin{equation}
H = \frac{J}{2} \sum_{\langle i,j \rangle} |h_i-h_j|^2 + \epsilon |h_i-h_j|^4.
\label{hamiltonian2}
\end{equation}
Here $0<\epsilon<1$ is a non-linearity parameter which controls the intensity of 
Schwoebel effect. The additional quartic term in \eq{hamiltonian2}, as the results presented here shows,
can be crucial for relaxation after ion irradiation of metallic surfaces as it is responsible 
for an uphill current which results in the formation of sharp peaked 'groove' structures.  

The algorithm for the Monte Carlo simulation is following. A site $1 \leq i \leq N$ 
is chosen at random and is subjected to follow erosion process with probability $p$ or the 
diffusive process with probability $1-p$. If erosion process is chosen, the angle $\phi_i$ 
is computed and a particle is eroded with probability $Y(\phi_i)p_e$. On the other hand 
if diffusive process is chosen, $w_{i,f}$ is computed using the surface Hamiltonian 
\eq{hamiltonian2} and the new configuration is updated. Time $\it{t}$ is incremented
by one unit.

\section{Experimental Details}
\label{exp}
High purity (99.99$\%$) Tantalum foils were bombarded by 3keV Ar ions under UHV conditions. 
The angle of incidence for ion beam was 15$^\circ$ w.r.t surface normal and its flux was 
10$^{13}$ions/cm$^2$. Scanning Probe Microscopy (SPM) studies have been conducted on the 
surfaces by using Bruker (Nanoscope V) system in tapping mode.

\section{Results and discussion}
\label{result}
\begin{figure}
\centering
\includegraphics[width=0.7\textwidth]{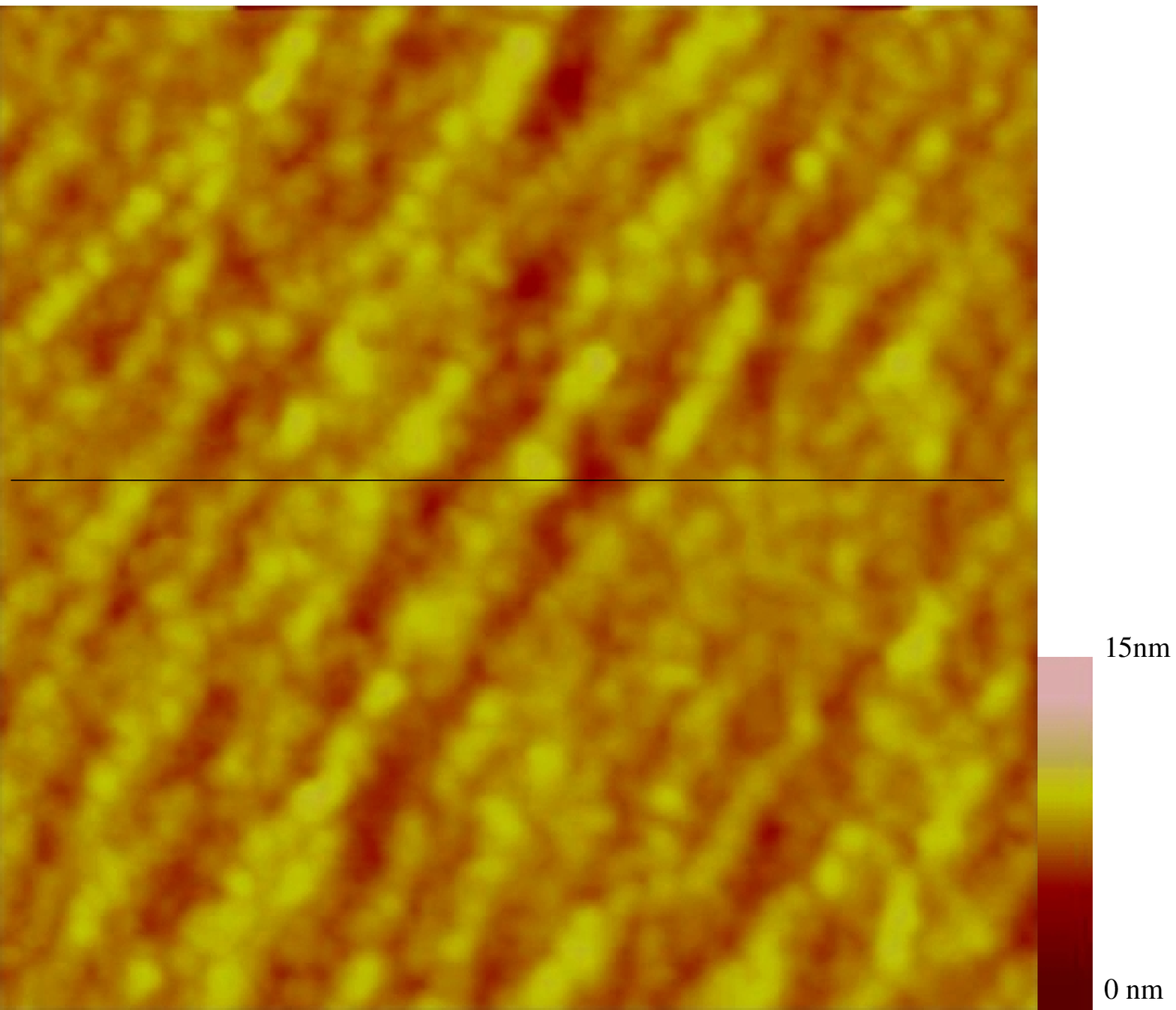}
\caption{Morphology of Tantalum surface: SPM image (500nm$\times$500nm)
after ion beam irradiation at a fluence of  36$\times10^{15}$ions/cm$^{2}$. }
\label{fig:afmimg}
\end{figure}

\begin{figure}
\centering
\includegraphics[width=0.7\textwidth,angle=-90 ]{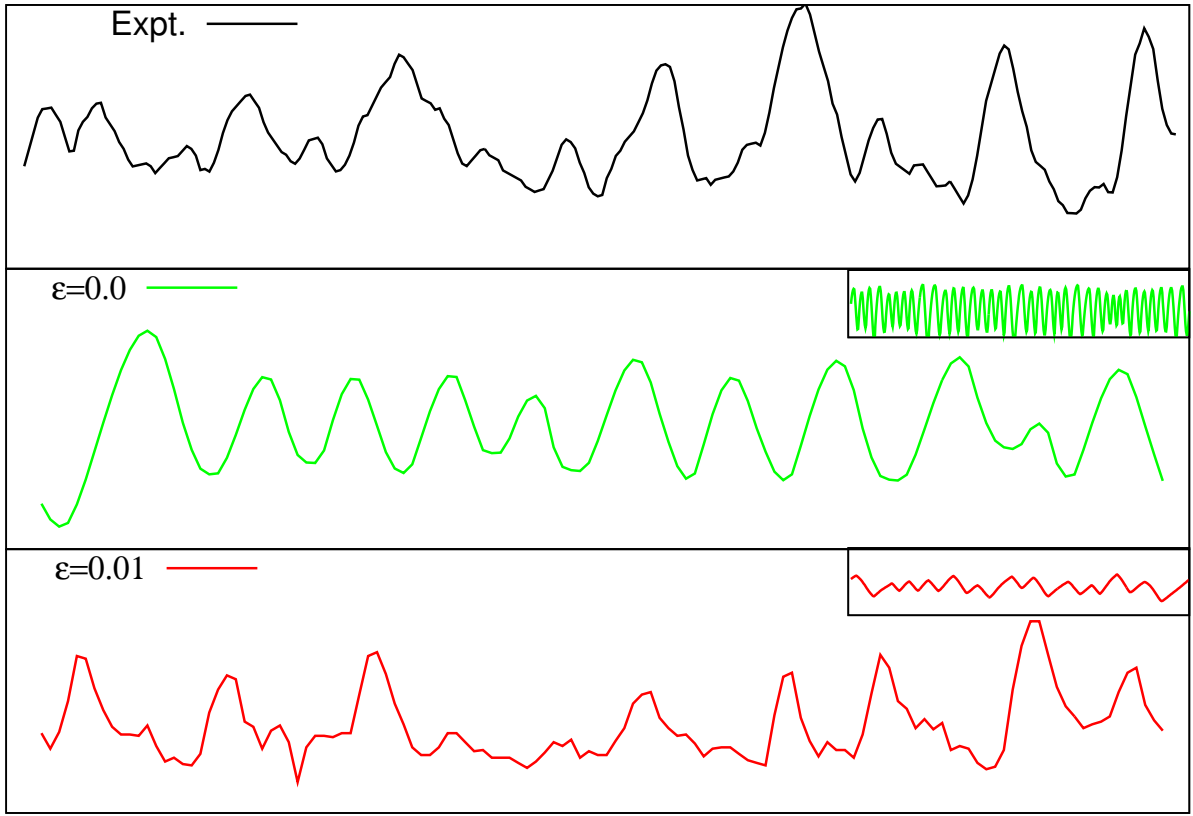}
\caption{Comparison of surface profile(h(x,t)) of experiment
with the simulations. The simulated profiles were obtained by KMC model with the parameters
$p=0.1$, $J_{c}/k_{B}T=0.25$ and $\epsilon$=0.0 or 0.01. Inset shows the
steady state profile of the surface
}
\label{fig:morph}
\end{figure}

\fig{fig:afmimg} displays an SPM image from a Tantalum surface bombarded by Ar$^+$ ions at fluence of 
3.6$\times10^{16}$ions/cm$^{2}$. A quasi periodic ripple pattern with a wavelength of $\sim$80nm is observed. 
1-dimension height profile from the experiment (section marked in \fig{fig:afmimg})
and from KMC simulations are presented in \fig{fig:morph}. 

For KMC simulations, first we study the model that considers erosion and includes relaxation 
mechanisms via only quadratic term in the Hamiltonian i.e for $\epsilon = 0.0$ in \eq{hamiltonian2}.
A range of parameters were chosen for simulations and the 
results are presented in \fig{fig:morph}, for $p=0.1$, $J_{c}/K_{B}T=0.25$. The height profile 
shows a periodic structure, usually similar to the morphologies observed 
for non metallic surfaces \cite{Jav} where the Schwoebel effects are not essential. 
Experimental height profile has several sharp peaks and grooves while the simulated profile
has only smooth morphology.

Next we examine the model where we consider relaxation of the surface by including both 
quadratic and quartic terms in the Hamiltonian \eq{hamiltonian2}. The non-linear parameter
$\epsilon$ is varied between 0.001 and 1.0. The surface morphology
with $\epsilon =0.01$ is presented in \fig{fig:morph} for $p=0.1$, $J_{c}/K_{B}T=0.25$. 
The simulated height profile here, with $\epsilon=0.01$,  displays good agreement
with experiments where formation of groove like structures are clearly observed. 
These features are characteristic signature of positive Schwoebel barrier that 
force the atom to move in uphill direction by breaking the translational invariance 
symmetry. A high diffusion rate, as observed here ($p=0.1$), is expected for metallic surfaces \cite{Val}.
The steady state height profile for KMC simulations are also shown in \fig{fig:morph} (inset).
 
The scaling behaviours and related exponents have also been explored here to investigate
the nature of the growing surface. The exponents are useful as they depend on the growth condition 
and not on the microscopic details of the system. The correlation length $\xi$, which 
represents the typical wavelength of fluctuations on the growing surface, also 
characterizes the phenomenon of kinetic roughening. The width of the surface grows as 
$W(t) \sim \xi(t)^{\alpha}$ for roughness exponent ${\alpha}$. The scale invariant surfaces
lead to scaling laws for correlation functions. The equal time height-height correlation (HHC)
function can be written as: 
\begin{equation}
 G(\mathbf{r},t)=L^{-1}\sum_{\mathbf{r'}}\langle[h(\mathbf{r+r'},t)-h(\mathbf{r'},t)]^{2}\rangle.
\label{hhc.eqn}
\end{equation}
Here $\mathbf{r}$ is the Translational length along lateral direction of the 1-d 
lattice and $\langle\cdot\rangle$ denotes the ensemble average. This HHC function has the 
following scaling form:
\begin{equation}
 G(\mathbf{r},t)= r^{2\alpha}g(r/\xi(t)).
\label{hhcscale.eqn}
\end{equation}
with $g(x)\sim\mbox{constant}$ for $ x \ll 1$ and $g(x)\sim x^{-2\alpha}$ for $x \gg 1$

\begin{figure}
\centering
\centerline{\includegraphics[width=0.7\textwidth, angle =-90 ]{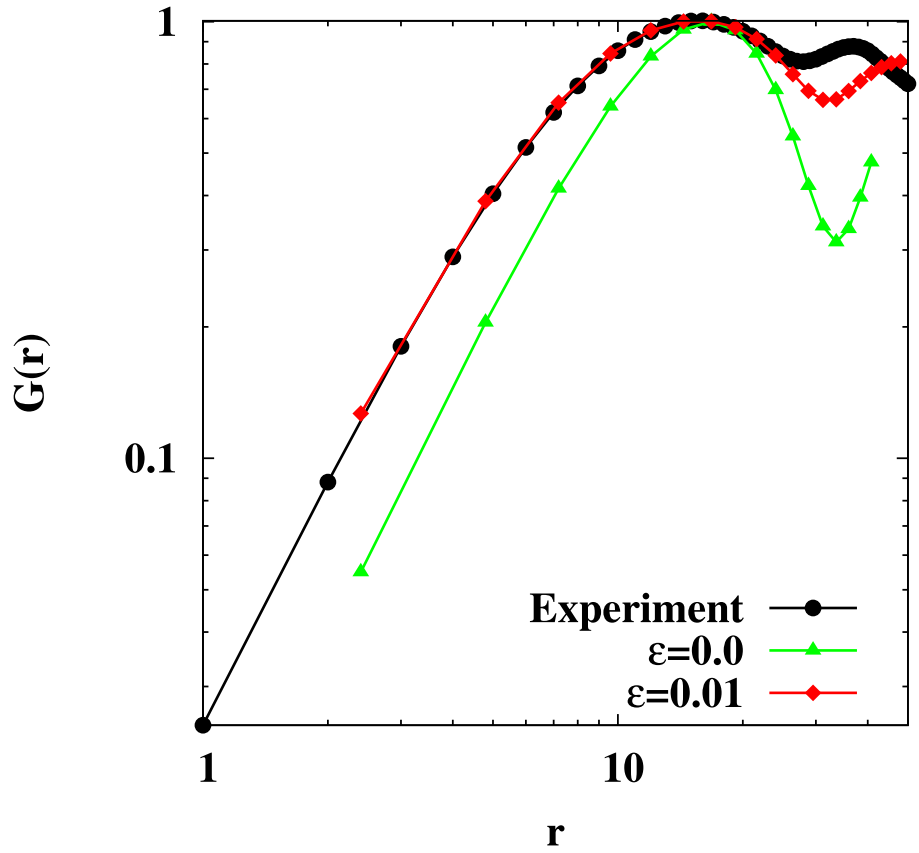}}
\caption{Equal time HHC \textbf{G(r,t)} as a function of translational distance \textbf{r}
for experiment. Simulation results are presented (for same parameters as in Fig.~2) with
$\epsilon$=0.0 or 0.01.  }
\label{fig:hhc}
\end{figure}

\fig{fig:hhc} presents the  1-dimensional height-height correlation function for experiment 
(using \fig{fig:afmimg}) as well as KMC simulations. By utilizing the phenomenological 
scaling function of form $H(r)\sim [1-\exp(-(r/\xi)^{2\alpha})]$, values of 
$\xi$ and $\alpha$ have been obtained and are listed in Table~1.
Although in absence of any Schwoebel effect ($\epsilon = 0.0$), the simulation results are very different from
experimental HHC function, after inclusion of Schwoebel effect ($\epsilon = 0.01$) the results are in 
agreement. These results demonstrate that a small nonlinearity 
parameter with $\epsilon =0.01$ is essential for achieving experimentally 
consistent HHC functional form, $\xi$ and $\alpha$. This
indicates that Schwoebel effect is necessary for understanding correct
growth behaviour on Tantalum surface. Value of 
$\alpha$ obtained here for KMC simulation, in absence of Schwoebel effect 
($\epsilon =0.0$), is similar to that observed in literature for MBE models
with linear Hamiltonian \cite{Wol,Das}. 

\begin{table}[!h]
\centering\smallskip\small\addtolength{\tabcolsep}{2pt}
\begin {tabular}{|c|c|c|}
\hline
                & $\alpha$       & $\xi$        \\ \hline
Experiment      & 1.22$\pm$0.26  & 5.66$\pm$0.36\\ \hline
$\epsilon=0.0$  & 1.63$\pm$0.35  & 2.63$\pm$0.50\\ \hline
$\epsilon=0.01$ & 1.20$\pm$0.07  & 7.14$\pm$0.05\\ \hline
\end{tabular}

\caption{Roughness exponent $\alpha$ and correlation length $\xi$.}
\label{tabl}
\end{table}

Obtaining $\alpha$ from G(r,t) can be difficult when the correlation length reaches the system size, 
specially in the steady state regime. In order to neglect finite size effects $\alpha$ can be 
computed  in $0\le r \le L/2$ regime, where $\it{L}$ is the system size. Structure Factor mentioned below
does not encounter this problem. The Structure Factor can be defined as \cite{Sie}:
\begin{equation}
S(\mathbf{k},t)=\langle\hat{h}(\mathbf{k},t)\hat{h}(-\mathbf{k},t)\rangle.
\end{equation}
Here, 
 $\hat{h}(\mathbf{r},t)=L^{-d/2}\sum_{\mathbf{r}}[h(\mathbf{r,t})-\overline{h}]e^{\mathbf{ikr}}$,
is the associated correlation function and $\overline{h}$ is the spatial average of h(r,t). 
This function has the following scaling form:
\begin{equation}
S(\mathbf{k},t)={k}^{-\gamma} s(\mathbf{k}^{zt}).
\end{equation}
with $\gamma = 2\alpha +d$. The scaling function {\it s} approaches a constant for large argument 
but behaves differently in the short time limit $x \ll 1$ where it has form
\begin{equation}
 s(x) \sim \left\{\begin{array}{c  l}
                      x \mbox{~if~}  \gamma\le z \\                                                                                                      
                      x^{\gamma/z} \mbox{~if~} \gamma \ge z                                                                                              
                  \end{array}\right.
\end{equation}

\begin{figure}
\centering
\includegraphics[width=0.7\textwidth, angle =-90 ]{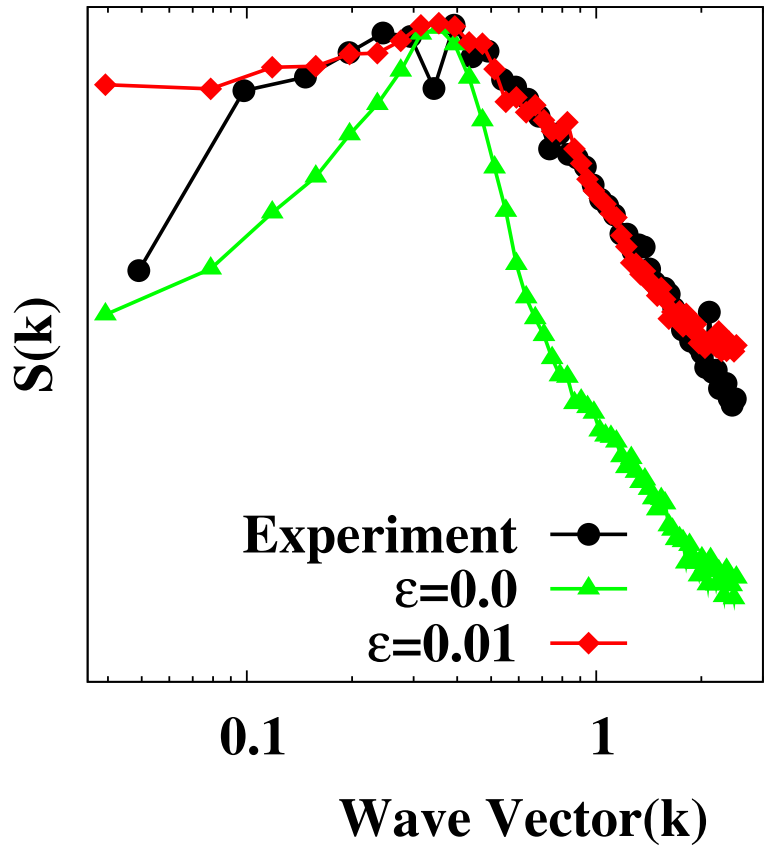}
\caption{Steady state structure factor \textbf{S(k)} for experiment. KMC simulation
results are presented (for same parameters as in Fig.~2) with $\epsilon$=0.0 or 0.01.}
\label{fig:struct}
\end{figure}

  In \fig{fig:struct}, the steady state structure factor 
$S(\mathbf {k})=S(\mathbf{k},t \rightarrow \infty)$ is shown for experiment as well as from KMC simulations
The result clearly demonstrate that the non-linear
Hamiltonian, with $\epsilon= 0.01$, agrees quite well with the experimental results.
For $\epsilon =0.0$, we observe that $S(\mathbf{k})$ qualitatively differs from the experimental results.
The value of exponent $\gamma$ obtained by linear Hamiltonian is 4.00$\pm$0.12 while for non-linear
Hamiltonian, the value is 3.27$\pm$0.07. For the ion beam modified Tantalum surfaces, $\gamma =3.04\pm0.06$ has 
been observed here. This value of $\gamma$ is not expected from
the universality classes of EW or KPZ. Thus the results indicate that positive Schwoebel
effect is necessary for understanding the morphological growth of ion sputtered Tantalum
surfaces which does not belong to the well known EW or the KPZ class.

\section{Conclusion}
In Conclusion, we have presented a KMC model to describe the morphology of ion sputtered
Tantalum surfaces. We have shown that a positive Schwoebel effect is needed to explain the
characteristics of self organized nanostructures observed in the experiment. The Schwoebel effect
has been used earlier for MBE growth models. Here, we have shown that, it can also be used for
ion sputtered metallic surfaces such as Tantalum. The scaling exponents computed from the simulation
and experimental data agrees quite well and indicates the presence of universality class that differs
from that of non-metallic surfaces.

\clearpage

\clearpage

\clearpage

\clearpage
\end{document}